\documentclass[12pt,epsf,fleqn]{article}
\usepackage{epsfig}
\usepackage[dvips]{color}
\begin{document}
\title{Study of Higgs self couplings of a supersymmetric $E_6$ model at the International Linear Collider}
\author{S. W. Ham$^{(1,2)}$\footnote{s.w.ham@hotmail.com},
Kideok Han$^{(1)}$\footnote{arini83@gmail.com},
Jungil Lee$^{(1,3)}$\footnote{jungil@korea.ac.kr},
and S. K. Oh$^{(4)}$\footnote{sunkun@konkuk.ac.kr}
\\
\\
{\it (1) Department of Physics, Korea University, Seoul 136-701, Korea} \\
{\it (2) School of Physics, KIAS, Seoul 130-722, Korea} \\
{\it (3) Korea Institute of Science and Technology Information} \\
{\it Daejeon, 305-806, Korea}  \\
{\it (4) Department of Physics, Konkuk University, Seoul 143-701, Korea} \\
\\
\\
}
\date{}
\maketitle
\begin{abstract}
We study the Higgs self couplings of a supersymmetric $E_6$ model that has two Higgs
doublets and two Higgs singlets.
The lightest scalar Higgs boson in the model may be heavier than 112 GeV, at the one-loop level,
where the negative results for the Higgs search at the LEP2 experiments are taken
into account.
The contributions from the top and scalar top quark loops are included in the
radiative corrections to the one-loop mass of the lightest scalar Higgs boson,
in the effective potential approximation.
The effect of the Higgs self couplings may be observed in the production of the lightest scalar Higgs bosons
in $e^+e^-$ collisions at the International Linear Collider (ILC) via double Higgs-strahlung process.
For the center of mass energy of 500 GeV with the integrated luminosity of 500 fb$^{-1}$
and the efficiency of 20 \%, we expect that at least 5 events of the lightest scalar Higgs boson may be
produced at the ILC via double Higgs-strahlung process.
\end{abstract}
\vfil\eject

%***********************************************************************
\section{Introduction}
%***********************************************************************

It is well accepted that one of the primary reasons to introduce the supersymmetry is
is to elucidate the hierarchy problem of the Standard Model (SM) [1-3].
In the SM, the existence of the neutral Higgs boson is generally explained within the context of
the origin of the electroweak symmetry breaking.
The breaking of the electroweak gauge symmetry is induced by
the self-interactions of the SM Higgs field, via the nontrivial vacuum expectation value.
Without supersymmetry, the mass of the SM Higgs boson would acquire quadratically
divergent loop corrections, requiring a cut-off scale at very high energy to confine
the Higgs mass at the electroweak scale.

If the supersymmetry is a symmetry of nature, the loop contributions of the particles may be
canceled by those of superpartners.
However, since experiments have never observed yet any superparticles that have the same mass as
quark or lepton, the supersymmetry cannot be exact.
In the presence of inexact supersymmetry, the one-loop corrections to the mass of the Higgs
boson is not so small but as large as 30 \% of the tree-level values,
due to incomplete cancelation.

There are a number of phenomenologically interesting supersymmetric models that embrace the SM.
A characteristic of these supersymmetric standard models is the requirement of
at least two Higgs doublets,
in order to give masses separately to the up-quark sector and the down-quark sector.
The key role of these Higgs fields is essentially the same as the SM Higgs field:
breaking of the electroweak gauge symmetry via self interactions.

The simplest one is the minimal supersymmetric standard model (MSSM) [4], which has just two Higgs doublets
and has been extensively studied.
The Higgs sector of the MSSM can be extended by introducing Higgs singlet fields [5,6].
In the next-to-minimal supersymmetric standard model (NMSSM), there is one additional Higgs singlet.
The motivation of introducing additional Higgs singlets is mainly to solve the $\mu$ problem of the MSSM [7].
In general, in terms of the vacuum expectation value of a neutral Higgs singlet, the $\mu$ parameter of
the MSSM might be generated dynamically.

Besides the MSSM and beyond, there are string-motivated supersymmetric models such as the supersymmetric $E_6$ model,
where the gauge symmetry is decomposed from $E_6$ to $SU(2) \times U(1) \times U_1 (1) \times U_2 (1)$.
The two extra $U(1)$ symmetries would be mixed to yield new linearly orthogonal combinations, $U'(1)$ and $U''(1)$.
Thus, it is a rank-6 supersymmetric model, with two extra neutral gauge bosons.
Its Higgs sector has two Higgs doublets and two Higgs singlets.
Phenomenology of the Higgs sector of this model has been investigated by several authors [8,9,10,11].
Some of us have elsewhere studied the possibility of detecting its neutral Higgs bosons
at the Large Hadron Collider, assuming explicit CP violation in its Higgs sector [11].

The self interactions of the Higgs fields are induced by cubic and quartic couplings in the Higgs potential.
Thus, in order to examine the Higgs phenomenology of a model, it is important to know the nature of
the Higgs self interactions and study the possibility of measuring them.
This kind of works have been performed some years ago in the SM and in the MSSM by Djouadi {\it et al}. [12,13].

In this article, we study the Higgs self couplings of the supersymmetric $E_6$ model
that has been investigated in Refs. [8,9] which we will call the SUSY $E_6$ model hereafter.
We study the possibility of discovering neutral scalar Higgs bosons of the SUSY $E_6$ model
at the International Linear Collider (ILC).
We calculate the cross section of the neutral Higgs bosons produced from double Higgs-strahlung process
in $e^+e^-$ collisions.
We find that, for reasonable parameter values, the cubic Higgs coupling of the lightest scalar
Higgs boson is obtained to be smaller than the corresponding coupling of the SM Higgs boson.
Thus, the production cross section of the lightest scalar Higgs boson via double Higgs-strahlung process
is smaller than that of the corresponding SM Higgs boson.
The radiative corrections to the mass of the lightest scalar Higgs boson due to the top and stop quark loops,
combined with the negative experimental result for the neutral scalar boson search at LEP2 [14],
suggest that the lightest scalar Higgs boson of the present model is heavier than 112 GeV.
At the International Linear Collider (ILC) with the center of mass energy of 500 GeV,
with the integrated luminosity of 500 fb$^{-1}$ and the efficiency of 20 \%,
we expect that at least 5 events of the double Higgs-strahlung process would be observed.

%***********************************************************************
\section{Higgs Potential}
%***********************************************************************

The SUSY $E_6$ model has two Higgs-doublet superfields, ${\cal H}_1$ and ${\cal H}_2$,
and two Higgs singlet superfields, ${\cal N}_1$ and ${\cal N}_2$.
The superpotential of the model is given by [8.9]
\begin{equation}
{\cal W} \approx h_t Q^T {\cal H}_2 t_R^c + \lambda {\cal H}_1 {\cal H}_2 {\cal N}_1  \ ,
\end{equation}
where $t_R^c$ is the right-handed top-quark superfield,
$Q$ is the left-handed quark doublet superfield of the third generation,
$h_t$ is the dimensionless Yukawa coupling coefficient for top quark, and $\lambda$ is a dimensionless coupling constant.
The bottom-quark superfield is absent as we take only the top-quark superfield.
Note that ${\cal N}_2$ is absent in the superpotential because the gauge symmetry of the model
prohibits its coupling to the Higgs doublet superfields.

The Higgs fields of the model are two Higgs doublets,
$H_1 = (H_1^0, H^-)$ and $H_2 = (H^+, H_2^0)$, and two Higgs singlets, $N_1$ and $N_2$.
The $\mu$ parameter of the MSSM is dynamically generated in this model in terms of
the vacuum expectation value of $N_1$.
In terms of these Higgs fields, the Higgs potential at the tree level can be written as [9]
\begin{eqnarray}
V_0 & = & m_1^2 |H_1|^2 + m_2^2 |H_2|^2 + m_3^2 |N_1|^2
+ m_4^2 |N_2|^2 - (\lambda A H_1 H_2 N_1 + {\rm H.c.} )  \cr
&  &\mbox{} + |\lambda|^2 [|H_1|^2 |H_2|^2 + |H_1|^2 |N_1|^2
+ |H_2|^2 |N_1|^2 ]  \  \cr
&  &\mbox{} + \left ( {g_2^2 \over 2} - |\lambda|^2 \right )  |H_1^{\dagger} H_2|^2
 + {g_1^2 + g_2^2 \over 8} (|H_1|^2 - |H_2|^2)^2  \cr
& &\mbox{} + {g_1^{'2} \over 72} [ C_{\theta} (|H_1|^2 + 4 |H_2|^2
- 5 |N_1|^2 - 5 |N_2|^2 ) \cr
& &\mbox{} - \sqrt{15} S_{\theta} (|H_1|^2
- |N_1|^2 + |N_2|^2  ) ]^2  \cr
& &\mbox{} + {g_1^{''2} \over 72} [ S_{\theta} (|H_1|^2 + 4 |H_2|^2
- 5 |N_1|^2 - 5 |N_2|^2 ) \cr
& & \mbox{} + \sqrt{15} C_{\theta} (|H_1|^2
- |N_1|^2 + |N_2|^2  ) ]^2   \   ,
\end{eqnarray}
where $m_i$ ($i=$ 1, 2, 3, 4) are the mass parameters, $g_1^{'}$ and $g_1^{''}$ are
respectively the gauge coupling coefficients of $U(1)'$ and $U(1)''$, respectively,
$A$ is the trilinear soft SUSY-breaking parameter with mass dimension,
$C_{\theta} = \cos \theta$, and $S_{\theta} = \sin \theta$,
with $\theta$ being the mixing angle between $U(1)'$ and $U(1)''$.

Among the parameters in the Higgs potential, $m_i$ may be expressed in terms of other parameters
by using the minimum equations that are obtained as the first derivatives of the Higgs potential
with respect to the four neutral Higgs fields.
After spontaneous breakdown of the electroweak gauge symmetry,
we have two additional gauge bosons $Z^{'}$ and $Z^{''}$ for the extra $U(1)'$ and $U(1)''$, respectively,
and seven physical Higgs particles: four scalar Higgs bosons, one pseudoscalar Higgs bosons,
and a pair of charged Higgs bosons.
The vacuum expectation values of the neutral Higgs fields are given as
$v_1 = \langle H_1\rangle $, $v_2 = \langle H_2\rangle $, $x_1 = \langle N_1\rangle $,
and $x_2 = \langle N_2\rangle $.
We introduce $\tan\beta = v_2/v_1$, and $v = \sqrt{v^2_1 +v^2_2} = 175$ GeV.

%************************************************************
\section{Mass spectra}
%************************************************************

Now, let us calculate the masses of the relevant particles.

%************************************************************
\subsection{Masses of neutral gauge bosons and scalar top quarks}
%************************************************************

In the SUSY $E_6$ model, there are three neutral gauge bosons, $Z$, $Z'$, and $Z''$
arising from $U(1)$, $U(1)'$, and $U(1)''$, respectively.
Their square masses are given as the eigenvalues of a symmetric $3\times 3$ matrix $M^G$,
the squared matrix.
The elements of $M^G$ are given explicitly as
\begin{eqnarray}
M_{11}^G &=& m_{Z}^2 \ , \cr
& & \cr
M_{22}^G &=& {20\over9} g_1^{'2} x_1^2 S_\theta^2 + {5\over36} g_1^{'2} x_2^2
  \bigg[ 8+7 C_{2\theta}-\sqrt{15} S_{2\theta} \bigg]  \cr
  & & \cr
  && + {1\over9} g_1^{'2} v^2
     \bigg[4-C_{2\theta}+\sqrt{15}\cos(2\beta) S_{2\theta} \bigg] \ , \cr
     & & \cr
M_{33}^G &=& {20\over9} g_1^{''2} x_1^2 C_{\theta}^2
  + {5\over36} g_1^{''2} x_2^2 \bigg[C_{\theta}+\sqrt{15}S_\theta \bigg]^2 \cr
  & & \cr
  && + {1\over9} g_1^{''2} v^2
     \bigg[4+C_{2\theta}-\sqrt{15}\cos(2\beta) S_{2\theta} \bigg] \ , \cr
     & & \cr
M_{12}^G &=& {1\over3} g_1^{'} m_{Z} v
  \bigg[\sqrt{3}C_{\theta} + \sqrt{5}\cos(2\beta) S_{\theta} \bigg] \ , \cr
  & & \cr
M_{13}^G &=& {1\over3} g_1^{''} m_{Z} v
  \bigg[\sqrt{5}\cos(2\beta)C_{\theta} - \sqrt{3} S_{\theta} \bigg] \ , \cr
  & & \cr
M_{23}^G &=& {10\over9} g_1^{'} g_1^{''} x_1^2  S_{2\theta}
  +{1\over9} g_1^{'} g_1^{''} v^2
    \bigg[\sqrt{15}\cos(2\beta)C_{2\theta}+S_{2\theta}\bigg] \cr
    & & \cr
  &&   - {5\over36} g_1^{'} g_1^{''} x_2^2
    \bigg[\sqrt{15} C_{2\theta}+7 S_{2\theta}\bigg] \ ,
\end{eqnarray}
where $m^2_Z = v^2(g^2_1 +g^2_2)/2$, $g_1$ and $g_2$ are
the gauge coupling coefficients of $U(1)$ and $SU(2)$, respectively,
$S_{2 \theta}= \sin (2 \theta)$ and $C_{2 \theta}= \cos (2 \theta)$.

A $3\times 3$ orthogonal matrix $U^G$ can be employed to diagonalize $M^G$,
in order to obtain three eigenvalues, namely, $m^2_Z$, $m^2_{Z'}$, and $m^2_{Z''}$.
We note that the smallest eigenvalue, $m^2_Z$, is different from the squared mass
of the neutral gauge boson in the SM.
However, in the decoupling limit of the $U(1)$ symmetry from the extra $U(1)'$ and $U(1)''$,
$m_Z$ becomes the gauge boson mass of the SM.
We sort them such that $m_Z < m_{Z'} < m_{Z''}$, where the mass of the lightest neutral gauge boson
should be equal to the mass of the existing neutral gauge boson, 91.2 GeV.

The mixings among them can be parameterized by three mixing angles, namely,
$\alpha_1$ between $Z$ and $Z'$,
$\alpha_2$ between $Z'$ and $Z''$, and
$\alpha_3$ between $Z$ and $Z''$, respectively.
These mixing angles may be expressed in terms of the elements of $U^G$ as
\begin{equation}
\alpha_1 = \arctan \left( {U_{12}^G \over U_{22}^G } \right) \ , \quad
\alpha_2 = \arcsin \left( -U_{32}^G \right)                  \ , \quad
\alpha_3 = \arctan \left( -{U_{31}^G \over U_{33}^G } \right) \ .
\end{equation}

We note that there are strong experimental constraints on the masses of the extra neutral gauge bosons
and on the mixing angles between the extra gauge bosons and the SM $Z$ boson.
The results of direct search in $p{\bar p}$ collisions at the Fermilab Tevatron
constrain that $m_{Z'}, \ m_{Z''} > 800$ GeV,
and the electroweak precision measurements at LEP2 constrain that
$|\alpha_1|$, $|\alpha_3| < 3 \times 10^{-3}$ [15].
Since the mixings between $U(1)$ and $U(1)'$ or between $U(1)$ and $U(1)''$ are very small,
the extra gauge symmetries are more or less decoupled from $SU(2) \times U(1)$.

Next, we calculate the scalar top quark masses at the tree level.
They are obtained as
\begin{equation}
m_{{\tilde t}_1, \ {\tilde t}_2}^2 =  {1 \over 2} (m_Q^2 + m_T^2) + m_t^2
+ {1 \over 4} m_Z^2 \cos 2 \beta + G_t^{'}
\mp \sqrt{X_t} \ ,
\end{equation}
with
\begin{eqnarray}
G_t^{'} & = &\mbox{} - {g_1^{' 2} \over 4} \left ( {1 \over 3} \sqrt{5 \over 2} S_{\theta} - {1 \over \sqrt{6} } C_{\theta} \right )
\left [ \left ( {\sqrt{10} \over 3} S_{\theta} + \sqrt{2 \over 3} C_{\theta} \right ) v^2 \cos^2 \beta
- {2 \over 3} \sqrt{10} S_{\theta} x_1^2  \right. \cr
& &\left. \mbox{} + \left ( {\sqrt{10} \over 3} S_{\theta} - \sqrt{2 \over 3} C_{\theta} \right ) v^2 \sin^2 \beta
- \left ( {1 \over 3} \sqrt{5 \over 2} S_{\theta} - {5 \over \sqrt{6}} C_{\theta} \right ) x_2^2 \right ]  \cr
& &\mbox{} - {g_1^{'' 2} \over 4} \left ( {1 \over 3} \sqrt{5 \over 2} C_{\theta} + {1 \over \sqrt{6} } S_{\theta} \right )
\left [ \left ( {\sqrt{10} \over 3} C_{\theta} - \sqrt{2 \over 3} S_{\theta} \right ) v^2 \cos^2 \beta
- {2 \over 3} \sqrt{10} C_{\theta} x_1^2  \right. \cr
& &\left. \mbox{} + \left ( {\sqrt{10} \over 3} C_{\theta} + \sqrt{2 \over 3} S_{\theta} \right ) v^2 \sin^2 \beta
- \left ( {1 \over 3} \sqrt{5 \over 2} C_{\theta} + {5 \over \sqrt{6}} S_{\theta} \right ) x_2^2 \right ]    \  ,
\end{eqnarray}
\begin{equation}
X_t = \bigg[ {1 \over 2} (m_Q^2 - m_T^2) + \left ({2 \over 3} m_W^2 - {5 \over 12} m_Z^2 \right ) \cos 2 \beta  \bigg]^2
+ m_t^2 ( A_t - \lambda x_1 \cot \beta )^2 \ ,
\end{equation}
where $m_Q$ and $m_T$ are the soft SUSY breaking masses,
$m^2_W = v^2 g^2_2 /2$,
$m_t = h_t v_2$ is the top quark mass,
$A_t$ is the trilinear SUSY breaking parameter with mass dimension,
$G_t^{'}$ is the $D$-term contribution from the extra $U(1)'$ and $U(1)''$,
and $X_t$ is the mixing term between ${\tilde t}_1$ and ${\tilde t}_2$.

%************************************************************
\subsection{Masses of pseudoscalar and scalar Higgs bosons}
%************************************************************

Next, we evaluate the one-loop corrections to the mass spectra.
The corrections are calculated by inserting the relevant tree-level masses
into the one-loop effective potential {\it a la} Coleman and Weinberg [16].
The top quark and scalar top contributions to the one-loop effective potential is given by
\begin{equation}
    V_1 = \sum_{l} {n_l {\cal M}_l^4 \over 64 \pi^2}
    \left [ \log {{\cal M}_l^2 \over \Lambda^2} - {3 \over 2} \right ]  \ ,
\end{equation}
where $\Lambda$ is the renormalization scale of the modified minimal substraction scheme.
The summation is over $l= {\tilde t}_1$, ${\tilde t}_2$, $t$ and $n_l = 6$
for the scalar top quarks and $n_l=-12$ for the top quark.

There are four neutral Higgs fields in the SUSY $E_6$ model, arising from two Higgs doublets and two Higgs singlets.
Among four complex components of these neutral Higgs fields, three of them are
gauged to generate masses of $Z$, $Z'$, and $Z''$;
and the remaining one complex component becomes the physical pseudoscalar Higgs boson of the model.
The mass of the pseudoscalar Higgs boson at the one-loop level is given as
\begin{equation}
    m_A^2 = m_{A^0}^2 + m_{A^1}^2   \ ,
\end{equation}
with
\begin{equation}
    m_{A^0}^2 = {2 \lambda A v  \over \sin 2 \alpha} \ , \quad
    m_{A^1}^2 = - {3 \lambda m_t^2 A_t  \over 8 \pi^2 v \sin 2 \alpha  \sin^2 \beta}
    f (m_{{\tilde t}_1}^2,  \ m_{{\tilde t}_2}^2)   \ ,
\end{equation}
where $m^2_{A^0}$ is obtained from the tree-level Higgs potential $V_0$,
$m^2_{A^1}$ is obtained from the one-loop Higgs potential $V_1$,
$\tan \alpha = v\sin 2 \beta/(2 x_1)$, and
\begin{equation}
 f(m_x^2, \ m_y^2) = {1 \over (m_y^2 - m_x^2)} \left[  m_x^2 \log {m_x^2 \over \Lambda^2} - m_y^2
\log {m_y^2 \over \Lambda^2} \right] + 1 \ .
\end{equation}

The second derivatives of the Higgs potential with respect to the real components
of these four neutral Higgs fields give us the mass matrix for four scalar Higgs bosons.
The $4\times 4$ symmetric squared mass matrix $M$ for the scalar Higgs bosons can be decomposed as
\begin{equation}
    M = M^0 + M^1   \   ,
\end{equation}
where $M^0$ is obtained from the tree-level Higgs potential $V_0$,
and $M^1$ from the one-loop Higgs potential $V_1$.
The explicit formulas for $M^0_{ij}$ ($i$, $j =$ 1, 2, 3, 4) are obtained as
\begin{eqnarray}
M_{11}^{0} & = & m_Z^2 \cos^2 \beta + m_{A^0}^2 \sin^2 \beta \cos^2 \alpha
+ {1 \over 18} (g_1^{'2} C_{\theta}^2 + g_1^{''2}S_{\theta}^2 ) v^2 \cos^2 \beta   \cr
& &\mbox{} + {5 \over 6} (g_1^{'2} S_{\theta}^2 + g_1^{''2}C_{\theta}^2) v^2 \cos^2 \beta
- {\sqrt{15} \over 9} (g_1^{'2} - g_1^{''2}) C_{\theta} S_{\theta} v^2 \cos^2 \beta   \ ,  \cr
M_{22}^{0} & = & m_Z^2 \sin^2 \beta + m_{A^0}^2 \cos^2 \beta \cos^2 \alpha
+ {8 \over 9} (g_1^{'2} C_{\theta}^2 + g_1^{''2}S_{\theta}^2 ) v^2 \sin^2 \beta \ ,  \cr
M_{33}^{0} & = & m_{A^0}^2 \sin^2 \alpha
+ {25 \over 18} (g_1^{'2} C_{\theta}^2 + g_1^{''2}S_{\theta}^2 ) x_1^2
+ {5 \over 6} (g_1^{'2} S_{\theta}^2 + g_1^{''2}C_{\theta}^2) x_1^2   \cr
& &\mbox{} - {5 \sqrt{15} \over 9} (g_1^{'2} - g_1^{''2}) C_{\theta} S_{\theta} x_1^2   \ ,  \cr
M_{44}^{0} & = &  {25 \over 18} (g_1^{'2} C_{\theta}^2 + g_1^{''2}S_{\theta}^2 ) x_2^2
+ {5 \over 6} (g_1^{'2} S_{\theta}^2 + g_1^{''2}C_{\theta}^2) x_2^2
+ {5 \sqrt{15} \over 9} (g_1^{'2} - g_1^{''2}) C_{\theta} S_{\theta} x_2^2   \ ,  \cr
M_{12}^{0} & = & (\lambda^2 v^2 - m_Z^2/2) \sin 2 \beta - m_{A^0}^2 \cos \beta \sin \beta \cos^2 \alpha   \cr
& &\mbox{}+ {1 \over 9} (g_1^{'2} C_{\theta}^2 + g_1^{''2}S_{\theta}^2 ) v^2 \sin 2 \beta
- {\sqrt{15} \over 9} (g_1^{'2} - g_1^{''2}) C_{\theta} S_{\theta} v^2 \sin 2 \beta    \ ,  \cr
M_{13}^{0} & = & 2 \lambda^2 v x_1 \cos \beta - m_{A^0}^2 \sin \beta \cos \alpha \sin \alpha
- {5 \over 18} (g_1^{'2} C_{\theta}^2 + g_1^{''2}S_{\theta}^2 ) v x_1 \cos \beta \cr
& &\mbox{}- {5 \over 6} (g_1^{'2} S_{\theta}^2 + g_1^{''2}C_{\theta}^2) v x_1 \cos \beta
+ {\sqrt{15} \over 3} (g_1^{'2} - g_1^{''2}) C_{\theta} S_{\theta} v x_1 \cos \beta    \ ,  \cr
M_{14}^{0} & = &\mbox{} - {5 \over 18} (g_1^{'2} C_{\theta}^2 + g_1^{''2}S_{\theta}^2 ) v x_2 \cos \beta
+ {5 \over 6} (g_1^{'2} S_{\theta}^2 + g_1^{''2}C_{\theta}^2) v x_2 \cos \beta \cr
& &\mbox{} + {2 \sqrt{15} \over 3} (g_1^{'2} - g_1^{''2}) C_{\theta} S_{\theta} v x_2 \cos \beta    \ ,  \cr
M_{23}^{0} & = & 2 \lambda^2 v x_1 \sin \beta - m_{A^0}^2 \cos \beta \cos \alpha \sin \alpha
- {10 \over 9} (g_1^{'2} C_{\theta}^2 + g_1^{''2}S_{\theta}^2 ) v x_1 \sin \beta \cr
& &\mbox{} + {2 \sqrt{15} \over 9} (g_1^{'2} - g_1^{''2}) C_{\theta} S_{\theta} v x_1 \sin \beta     \ ,  \cr
M_{24}^{0} & = &\mbox{} - {10 \over 9} (g_1^{'2} C_{\theta}^2 + g_1^{''2}S_{\theta}^2 ) v x_2 \sin \beta
- {2 \sqrt{15} \over 9} (g_1^{'2} - g_1^{''2}) C_{\theta} S_{\theta} v x_2 \sin \beta       \ ,  \cr
M_{34}^{0} & = & {25 \over 18} (g_1^{'2} C_{\theta}^2 + g_1^{''2}S_{\theta}^2 ) x_1 x_2
- {5 \over 6} (g_1^{'2} S_{\theta}^2 + g_1^{''2}C_{\theta}^2) x_1 x_2 \ .
\end{eqnarray}
and the explicit formulae for $M^1_{ij}$ ($i,j =$ 1,2,3,4) are obtained
by inserting the Higgs field dependant masses of the top quark and scalar top quarks at the tree-level
into $V^1$ as
\begin{eqnarray}
M_{ij}^1 & = & {3 \over 32 \pi^2 v^2} W_i W_j
{g(m_{{\tilde t}_1}^2, m_{{\tilde t}_2}^2) \over (m_{{\tilde t}_2}^2 - m_{{\tilde t}_1}^2)^2}
+ {3 \over 32 \pi^2 v^2} A_i A_j
\log \left ( {m_{{\tilde t}_1}^2 m_{{\tilde t}_2}^2 \over \Lambda^4 } \right ) \cr
& &\mbox{} + {3 \over 32 \pi^2 v^2} (W_i A_j + A_i W_j)
{ \log ( m_{{\tilde t}_2}^2/ m_{{\tilde t}_1}^2)  \over (m_{{\tilde t}_2}^2 - m_{{\tilde t}_1}^2)} + D_{ij} \ ,
\end{eqnarray}
with
\begin{equation}
 g(m_x^2,m_y^2) = {m_y^2 + m_x^2 \over m_x^2 - m_y^2} \log {m_y^2 \over m_x^2} + 2 \ ,
\end{equation}
\begin{eqnarray}
W_1 & = & {2 m_t^2 x_1 \lambda \Delta_1 \over \sin \beta } + \cos \beta \Delta_2 \ , \cr
W_2 & = & -{2 m_t^2 A_t \Delta_1 \over \sin \beta } - \sin \beta \Delta_2   \ , \cr
W_3 & = & {2 m_t^2 \lambda v \Delta_1 \over \tan \beta}  \ , \cr
W_4 & = & 0 \ , \cr
A_1 & = & {1\over 2} \cos \beta (4 G_1 v^2 + m_Z^2) \ , \cr
A_2 & = & {2 m_t^2 \over \sin \beta} + 2 G_2 v^2 \sin\beta - {m_Z^2 \over 2} \sin \beta  \ , \cr
A_3 & = & 2 G_3 x_1 v  \ ,  \cr
A_4 & = & 2 G_4 x_2 v  \ ,
\end{eqnarray}
\begin{eqnarray}
\Delta_1 & = & \lambda x_1 \cot \beta - A_t \ , \cr
\Delta_2 & = & \bigg ({4 \over 3} m_W^2 - {5 \over 6} m_Z^2 \bigg )
\bigg (m_Q^2 - m_T^2 + \bigg ({4 \over 3} m_W^2 - {5 \over 6} m_Z^2 \bigg) \cos 2 \beta \bigg)   \  ,
\end{eqnarray}
\begin{eqnarray}
G_1 & = & {g_1^{' 2} \over 36} (4 C_{2 \theta} - 1) - {g_1^{''2} \over 36} (4 C_{2 \theta} + 1) \ , \cr
& & \cr
G_2 & = & {g_1^{' 2} \over 36} (\sqrt{15} S_{2 \theta} + C_{2 \theta} - 4)
- {g_1^{''2} \over 36} (\sqrt{15} S_{2 \theta} + C_{2 \theta} + 4)  \ , \cr
& & \cr
G_3 & = &\mbox{} - {g_1^{' 2} \over 18} (\sqrt{15} C_{\theta} - 5 S_{\theta} ) S_{\theta}
+ {g_1^{''2} \over 18} (\sqrt{15} S_{\theta} + 5 C_{\theta} ) C_{\theta} \ , \cr
& & \cr
G_4 & = & {g_1^{' 2} \over 72} (10 - 3 \sqrt{15} S_{2 \theta} + 5 C_{2 \theta})
+ {g_1^{''2} \over 72} (10 + 3 \sqrt{15} S_{2 \theta} - 5 C_{2 \theta})  \ ,
\end{eqnarray}
\begin{eqnarray}
D_{11} & = & m_{A^1}^2 \sin^2 \beta \cos^2 \alpha  - {3 \cos^2 \beta \over 16 \pi^2 v^2}
\left( {4 m_W^2 \over 3} - {5 m_Z^2 \over 6} \right)^2
f(m_{{\tilde t}_1}^2, \ m_{{\tilde t}_2}^2) \ , \cr
& & \cr
%%%%%%%%%
D_{22}
& = & m_{A^1}^2 \cos^2 \beta \cos^2 \alpha
- {3 \sin^2 \beta \over 16 \pi^2 v^2}
\left( {4 m_W^2 \over 3} - {5 m_Z^2 \over 6} \right)^2
f(m_{{\tilde t}_1}^2, \ m_{{\tilde t}_2}^2)     \cr
& & \cr
& &\mbox{} - {3 m_t^4 \over 4 \pi^2 v^2 \sin^2 \beta} \log \left ({m_t^2 \over \Lambda^2} \right )  \ , \cr
& & \cr
%%%%%%%%%
D_{33} & = &   m_{A^1}^2 \sin^2 \alpha \ , \cr
& & \cr
%%%%%%%%%
D_{12}
& = &\mbox{} {3 \sin 2 \beta \over 32 \pi^2 v^2} \left ({4 m_W^2 \over 3} - {5 m_Z^2 \over 6} \right)^2
f(m_{{\tilde t}_1}^2, \ m_{{\tilde t}_2}^2) \ , \cr
& & \cr
%%%%%%%%%
D_{13} & = &\mbox{} - m_{A^1}^2 \sin \beta \cos \alpha \sin \alpha
- {3 m_t^2 \lambda^2 x_1 \cos \beta \over 8 \pi^2 v \sin^2 \beta}
f(m_{{\tilde t}_1}^2, \ m_{{\tilde t}_2}^2)  \ , \cr
& &  \cr
%%%%%%%%%
D_{23}
& = &\mbox{} - m_{A^1}^2 \cos \beta \cos \alpha \sin \alpha \ ,  \cr
& & \cr
%%%%%%%%%
D_{i4} & = & 0 \ , ~~(i = 1,2,3,4)  \ .
\end{eqnarray}
%%%%%%%%%%%%%%%%%%%%%%%%%%%%%%%%%%%%
The eigenstates of $M$ are four physical scalar Higgs bosons of the SUSY $E_6$ model,
denoted as $S_i$ ($i =$ 1, 2, 3, 4), and the corresponding eigenvalues are their squared masses,
denoted as $m^2_{S_i}$.
The masses of these four scalar Higgs bosons are ordered as
$m_{S_1} < m_{S_2} < m_{S_3} < m_{S_4}$.

For the numerical analysis, we establish the parameter space as follows:
$1 < \tan \beta \leq 30$, $0 < \lambda \leq 0.83$, $ 10 < A < 400$,
$0  < \theta < \pi/2$, $100 \leq x_1, x_2 \leq 1500$ GeV,
$100 \leq m_Q, \ m_T, \ A_t \leq 1000$ GeV.
Furthermore, we put $m_t$ = 175 GeV, and we assume that the scalar top quarks are heavier than the top quark.
We set the lower bound on the effective $\mu^{\rm e} = \lambda x_1$ parameter to be 150 GeV,
in order to take into account the experimental lower bound on the lighter chargino mass at the LEP2 experiments.

We also take into account the experimental results for the Higgs search at the LEP2 experiments.
Recently, the collaborations of the LEP2 experiments have reported the model-independent upper bound
on the Higgs coupling coefficient to a pair of $Z$ bosons at 95 \% confidence level [14].
The result may be interpreted for the coefficient of the SUSY $E_6$ model.
The coupling coefficient of $S_i$ to a pair of $Z$ bosons of the model,
normalized by the corresponding SM coupling coefficient, can be written as
\begin{equation}
G_{ZZS_i} \approx \cos \beta O_{1i} + \sin \beta O_{2i} \ ,
\end{equation}
where $O_{ij}$ are the $ij$-th element of the $4\times 4$ orthogonal matrix that diagonalizes
the mass matrix for the neutral scalar Higgs bosons.

For a given set of parameter values, we evaluate $m_{S_i}$ and $G_{ZZS_i}$.
In this way, using Monte Carlo method, we explore $10^5$ points of the established parameter space.
In Fig. 1, we show the result of evaluations of $m_{S_1}$ and $G_{ZZS_1}$.
The SM coupling coefficient is 1.0 since it is also normalized by itself.
We find that $G_{ZZS_1}$ is larger than 0.9 for the most of the parameter space,
which tells us that $S_1$ is more or less equivalent to the SM Higgs boson.
Since $G_{ZZS_i}$ ($i =$ 1, 2, 3, 4) satisfy a sum rule of $\sum^4_{i =1} G_{ZZS_i}^2 \approx 1$,
this implies that the pair of $Z$ bosons couples nearly exclusively to $S_1$ in most of the parameter space
and the couplings to the other heavier scalar Higgs bosons are negligibly small.

The result in Fig. 1 also shows that $m_{S_1}$ is in the range of 112 to 142 GeV at the one-loop level.
The upper bound on $m_{S_1}$ is determined primarily by the maximum value of the SUSY breaking scale
and the LEP2 constraints on the SM Higgs coupling coefficient to a pair of $S$ bosons.
The masses of heavier scalar Higgs bosons are
135 GeV $< m_{S_2} <$ 897 GeV, 800 GeV $< m_{S_3} <$ 1155 GeV, and 1033 GeV $< m_{S_4} <$ 2828 GeV.
Also, the masses of the pseudoscalar Higgs boson of the SUSY $E_6$ model
at the one-loop level
are estimated to be $120 < m_A < 2828$ for the parameter
values we consider.

The coupling coefficient $G_{ZZS_1}$ is crucial to calculate the production cross section
for the Higgs-strahlung process, $\sigma(e^+e^- \rightarrow Z \rightarrow Z S_1)$.
In Fig. 2, we show the production cross section of $S_1$ via Higgs-strahlung process in
$e^+ e^-$ collisions with $\sqrt{s} = 500$ GeV,
where the corresponding SM cross section is also shown in a solid curve,
as a function of the SM Higgs boson mass.
The values of the relevant parameters are the same as in Fig. 1,
chosen randomly in the established parameter space.

One may notice in Fig. 2 that most of the $10^5$ points are distributed quite close to the solid curve.
In other words, $S_1$ of the SUSY $E_6$ model behaves almost the same as the SM Higgs boson in the
Higgs-strahlung process in $e^+e^-$ collisions.
One may also notice in Fig. 2 that the absolute minimum of the production cross section is about 19 fb.
That is, parameter values, the production cross section is larger than 19 fb for any.
It is worthwhile to compare this absolute lower bound on the production cross section with
the corresponding results of other models.
We have studied the same subject in the next-to MSSM (NMSSM) [17].
We have calculated that the absolute lower bound on the production cross section of
the lightest scalar Higgs boson via Higgs-strahlung process in the NMSSM is about 15 fb
(see the dashed curve of Fig. 3a in Ref. [17]).
Thus, the absolute lower bound on the production cross section of the SUSY $E_6$ model is
in fact larger than the corresponding value in the NMSSM.

It is usually anticipated that, for a supersymmetric model with several Higgs singlets,
the production cross section for the lightest scalar Higgs boson tends to decrease
as the number of Higgs singlets increases,
because the probability of production should be shared with other heavier scalar Higgs bosons.
In this respect, the lower bound on the production cross section of the SUSY $E_6$ model should be smaller
than that of the NMSSM,
since the SUSY $E_6$ model has one more Higgs singlet than the NMSSM.
However, the result is opposite to the usual anticipation.
We think that the main reason for this result is that $S_1$ of the SUSY $E_6$ model is more similar
to the SM Higgs boson than the lightest scalar
Higgs boson of the NMSSM, at least in the Higgs-strahlung process in $e^+e^-$ collisions.

%*****************************************************************
\section{Production of scalar Higgs bosons via double Higgs-strahlung process}
%*****************************************************************

The cubic and quartic couplings of Higgs bosons contribute when a process involves
multiple Higgs bosons.
Typically, the double Higgs production process in $e^+e^-$ collisions depends on
the cubic couplings of Higgs bosons and the quartic couplings of Higgs-Higgs-gauge-gauge bosons.
For practical simplicity, we only consider double $S_1$ productions: $e^+e^- \rightarrow ZS_1 S_1$.
The double productions of heavier scalar Higgs bosons are kinematically suppressed and
therefore the production cross sections are very small.
The relevant Feynman diagrams for the double $S_1$ production are shown in Fig. 3.
Notice that the quartic coupling and the cubic coupling are present in Fig. 3.

In principle, all of the four scalar Higgs bosons of the SUSY $E_6$ model may participate in Fig. 3
as the intermediate particle.
The contributions of the heavier scalar Higgs particles as the intermediate particle depend on two
coupling coefficients: $G_{ZZS_i}$ at the one end and $G_{S_i S_1 S_1}$ at the other end ($i =$ 2, 3, 4).
However, as we have observed before, $G_{ZZS_i}$ ($i =$ 2, 3, 4), the coupling coefficients of
heavier scalar Higgs bosons to the $Z$ boson pair, are negligibly smaller than $G_{ZZS_1}$.
Thus, the contributions of the heavier scalar Higgs particles as the intermediate particle may be neglected,
and we only consider $S_1$ as the intermediate particle in Fig. 3 in our numerical analysis.
Consequently, only $S_1$ participates in Fig. 3.

In order to calculate the production cross section of the lightest scalar Higgs boson,
we have to know its cubic and quartic coupling coefficients in the SUSY $E_6$ model.
The cubic coupling coefficient of $S_i$ can be written as
\begin{equation}
    G_{S_i S_i S_i} = G_{S_i S_i S_i}^0 + G_{S_i S_i S_i}^1
\end{equation}
where $G_{S_i S_i S_i}^0$ is the tree-level coefficient and $G_{S_i S_i S_i}^1$ is the one-loop
correction.
Explicitly, the tree-level cubic coupling coefficient of $S_1$ is given as
\begin{eqnarray}
G_{S_i S_i S_i}^0 &=& 12\lambda^2 [ (O_{1i}^2 + O_{2i}^2 ) O_{3i} x_1
      + O_{1i} (O_{2i}^2 + O_{3i}^2) v \cos\beta
      + O_{2i} (O_{1i}^2 + O_{3i}^2) v \sin\beta] \cr
&& + 3v(g_1^2+g_2^2) (O_{1i}^2 - O_{2i}^2) (O_{1i} \cos\beta -
O_{2i} \sin\beta)
       - \frac{6 m_{A^0}^2\sin2\alpha}{v} O_{1i}O_{2i}O_{3i}   \cr
&&  + G_{S_i S_i S_i}^{'} + G_{S_i S_i S_i}^{''}
\end{eqnarray}
with
\begin{eqnarray}
G_{S_i S_i S_i}^{'} &=& {1\over 3} g_1^{'2} [C_\theta (O_{1i}^2 + 4
O_{2i}^2 - 5 O_{3i}^2 - 5 O_{4i}^2) \cr
& &\mbox{}  - \sqrt{15} (O_{1i}^2 - O_{3i}^2 + O_{4i}^2 ) S_\theta]
      \times [\sqrt{15}
        S_\theta (O_{3i} x_1 - O_{4i} x_2 - O_{1i} v \cos\beta)   \cr
        &&  -  C_\theta (5 O_{3i} x_1 + 5 O_{4i} x_2 - O_{1i} v \cos\beta -
         4 O_{2i} v \sin\beta) ]    \    ,    \cr
G_{S_i S_i S_i}^{''} &=& G_{S_i S_i S_i}^{'} \bigg[g_1^{'} \to
g_1^{''} , C_\theta \to S_\theta, S_\theta \to -C_\theta \bigg] ,
\end{eqnarray}
where $G_{S_i S_i S_i}^{'}$ the $D$-term contribution of the $U(1)'$
and $G_{S_i S_i S_i}^{''}$ the $D$-term contribution of the $U(1)''$
at the tree level.

The one-loop correction to the cubic coupling of the scalar Higgs boson is given as
\begin{eqnarray}
G_{S_i S_i S_i}^1 & = &\mbox{} - {9 \over 32 \pi^2} Y_{ii} Y_i
\log \bigg ( {m_t^2 \over \Lambda^2} \bigg ) - {3 \over 32 \pi^2 m_t^2} Y_i^3  \cr
& &\mbox{} + {9 \over 64 \pi^2} A_{ii} A_i
\log \bigg ( { m_{{\tilde t}_1}^2 m_{{\tilde t}_2}^2 \over \Lambda^4 } \bigg )
- {9 \over 32 \pi^2} W_i^3 { \log (m_{{\tilde t}_1}^2 m_{{\tilde t}_2}^2/\Lambda^4) \over (m_{{\tilde t}_2}^2
- m_{{\tilde t}_1}^2)^4} \cr
& &\mbox{} + {9 \over 64 \pi^2} W_{ii} W_i
{ \log (m_{{\tilde t}_1}^2 m_{{\tilde t}_2}^2/\Lambda^4) \over (m_{{\tilde t}_2}^2 - m_{{\tilde t}_1}^2)^2} \cr
& &\mbox{} + {9 \over 64 \pi^2} \bigg [ A_{ii} W_i + W_{ii} A_i \bigg ]
{ \log (m_{{\tilde t}_2}^2 / m_{{\tilde t}_1}^2) \over (m_{{\tilde t}_2}^2 - m_{{\tilde t}_1}^2)
} - {9 \over 32 \pi^2} W_i^2 A_i
{ \log (m_{{\tilde t}_2}^2 / m_{{\tilde t}_1}^2) \over (m_{{\tilde t}_2}^2
- m_{{\tilde t}_1}^2)^3} \cr
& &\mbox{} + {3 \over 64 \pi^2 m_{{\tilde t}_1}^2}
\bigg [A_i - {W_i  \over (m_{{\tilde t}_2}^2 - m_{{\tilde t}_1}^2)} \bigg ]^3
+ {3 \over 64 \pi^2 m_{{\tilde t}_2}^2}
\bigg [A_i + {W_i \over (m_{{\tilde t}_2}^2 - m_{{\tilde t}_1}^2)} \bigg ]^3 \ ,
\end{eqnarray}
where
\begin{eqnarray}
Y_{i} &=& \frac{2 m_t^2 O_{2i}}{v\sin\beta} , \cr
Y_{ii} &=& \frac{2 m_t^2 O_{2i}^2}{v^2 \sin^2\beta} , \cr
A_{i} &=& 2 g_1^{'} O_{3i} x_1+2 g_1^{''} O_{4i} x_2+2 g_1 O_{1i}
v \cos\beta+2 g_2 O_{2i} v \sin\beta + \frac{2 m_t^2}{v\sin\beta} O_{2i}  \cr
&& +\frac{(O_{1i} \cos\beta-O_{2i} \sin\beta) m_Z^2}{2 v}, \cr
A_{ii} &=& 2 g_1 O_{1i}^2+2 g_2 O_{2i}^2 +
\frac{2m_t^2}{v^2\sin^2\beta}O_{2i}^2+2 g_1^{'} O_{3i}^2+2 g_1^{''}
O_{4i}^2+\frac{\left(O_{1i}^2-O_{2i}^2\right) m_Z^2}{2 v^2}  \ , \cr
W_{i} &=& \frac{O_{1i} \cos\beta-O_{2i} \sin\beta}
{v}\left(\frac{4 m_W^2}{3}-\frac{5 m_Z^2}{6}\right) \cr
  && \times \left[m_Q^2-m_T^2+\cos(2 \beta) \left(\frac{4 m_W^2}{3}-\frac{5 m_Z^2}{6}\right)\right] \cr
  && + \frac{2m_t^2}{v\sin\beta}(O_{1i} x_1 \lambda +O_{3i} v \lambda  \cos\beta - A_t O_{2i}) (x_1 \lambda  \cot\beta-A_t ), \cr
W_{ii} &=& \frac{2(O_{1i} \cos\beta-O_{2i}
\sin\beta)^2}{v^2}\left(\frac{4 m_W^2}{3}-\frac{5 m_Z^2}{6}\right)^2
\cr
  && +\frac{O_{1i}^2-O_{2i}^2}{ v^2}\left(\frac{4 m_W^2}{3}-\frac{5 m_Z^2}{6}\right) \left[m_Q^2-m_T^2+ \cos(2 \beta) \left(\frac{4 m_W^2}{3}-\frac{5 m_Z^2}{6}\right)\right] \cr
  && + 2 \left(\frac{m_t}{v\sin\beta}\right)^2 \bigg[O_{3i}^2 v^2 \lambda ^2 \cos^2\beta-2 O_{3i} v \lambda \{(A_t O_{2i}-2 O_{1i} x_1 \lambda ) \cos\beta \cr
  && +A_t O_{1i} \sin\beta\} + (A_t O_{2i}-O_{1i} x_1 \lambda )^2\bigg]   \  .
\end{eqnarray}

The corresponding SM cubic coupling coefficient, $G_{hhh}$, may be obtained from the SM
Higgs potential $\mu^2\phi^\dag \phi/2 +\lambda (\phi^\dag \phi)^2/4$ as $6 m^2_h /v$, where
$v = \langle \phi \rangle$ and $m_h$ is the mass of SM Higgs boson.
Conventionally, the SM cubic coupling coefficient is normalized such that [13]
${\bar G}_{hhh} = G_{hhh}\cdot v/(2m^2_Z)$.
In Fig. 3, we also plot ${\bar G}^2_{hhh}$ as a function of $m_h$ (the solid curve).

In order to compare the cubic coupling coefficient of the SUSY $E_6$ model with the SM cubic coupling coefficient,
we may normalize $G_{S_i S_i S_i}$ in the unit of $2 m_Z^2/v$ as
$[v/ (2 m_Z^2)] G_{S_i S_i S_i}$.
In Fig. 4 we plot the square of the cubic coupling coefficient, normalized by $2 m_Z^2/v$, against $m_{S_1}$,
by varying the values of the relevant parameters within the established space.
We find that the cubic coupling coefficient of the SUSY $E_6$ model is smaller than
the corresponding SM coefficient.
This behavior of the SUSY $E_6$ model is similar to the MSSM, where
the cubic coupling coefficient of the lightest scalar Higgs boson is smaller
than the corresponding coefficient of the SM Higgs boson [18].
On the other hand, it is different to the two Higgs doublet model, where
the SM cubic coupling coefficient may be smaller than the cubic coupling coefficient of
the lightest scalar Higgs boson [19].
We also find that the cubic coupling coefficients of other scalar Higgs bosons, $G_{S_i S_i S_i}$ ($i=$ 2,3,4)
are smaller than the SM cubic coupling coefficient.

Next, let us consider the quartic coupling coefficient of Higgs-Higgs-gauge-gauge bosons of the SUSY $E_6$ model.
When normalized by the corresponding SM quartic coupling coefficient,
$G_{ZZS_iS_i}$ of $S_i$ pair to $Z$ boson pair is given as
\begin{equation}
    G_{ZZS_iS_i} \approx O_{1i}^2 + O_{2i}^2 \ ,
\end{equation}
if the mixing between $Z$ boson and extra neutral gauge bosons is neglected.
Since the mixing angles $|\alpha_2|$ and $|\alpha_3|$ are very small, the neglection is practically
justified.

Now, we are ready to calculate the differential cross section for the double Higgs-strahlung process
in $e^+ e^-$ collisions of the lightest scalar Higgs boson of the SUSY $E_6$ model.
It is obtained as
\begin{equation}
    \frac{d \sigma }{d x_1 d x_2} (e^+e^- \rightarrow Z \rightarrow ZS_1 S_1) =
    \frac{\sqrt{2} G_F^3 m_Z^6}{384 \pi^3 s}
    \frac{v_e^2 + a_e^2}{(1- \mu_Z)^2}\, {\cal Z}
\end{equation}
where $x_i = 2 E_i /\sqrt{s}$ is the scaled energy of the $i$-th $S_1$,
with $E_i$ being the energy of the $i$-th $S_1$ and
$\sqrt{s}$ being the c.m. energy of $e^+e^-$ system ($i = $ 1,2),
$v_e = -1 + 4\sin^2 \theta_W$ and $a_e = -1$ are respectively the
vector and axial-vector $Z$ charges of the incoming electron,
$\mu_Z = m^2_Z /s$ is the square of the reduced $Z$ boson mass, and
${\cal Z}$ is given by
\begin{eqnarray}
{\cal Z} & = & {\cal Z}_1 G_{ZZS_1}^2 G_{S_1 S_1 S_1}^2 + {\cal Z}_2 G_{ZZS_1}^4
+ {\cal Z}_3 G_{ZZS_1 S_1}^2 \cr
& &\mbox{} + {\cal Z}_{12} G_{ZZS_1}^3 G_{S_1 S_1 S_1}
+ {\cal Z}_{13} G_{ZZS_1} G_{S_1 S_1 S_1} G_{ZZS_1S_1}
+ {\cal Z}_{23} G_{ZZS_1}^2 G_{ZZS_1S_1} \ ,
\end{eqnarray}
with
\begin{eqnarray}
{\cal Z}_1 &=& \mu_Z\frac{ (y_1+y_2)^2+8 \mu_Z}{4
(y_3-\mu_{HZ})^2} \ , \cr %%
{\cal Z}_2 &=& \mu_Z \frac{(y_1+y_2)^2+8 \mu_Z}{(y_1+\mu_{HZ})^2}
 +\mu_Z\frac{(y_1+y_2)^2+8 \mu_Z}{(y_2+\mu_{HZ})^2}
+ \frac{2 \mu_Z [(y_1+y_2)^2+8 \mu_Z]}{(y_1+\mu_{HZ}) (y_2+\mu_{HZ})}  \cr
  && -\frac{y_1 \mu_H (y_1-4 \mu_Z+y_1 \mu_Z)}{(y_1+\mu_{HZ})^2 \mu_Z}
   -\frac{y_2 \mu_H (y_2-4 \mu_Z+y_2 \mu_Z)}{\mu_Z(y_2+\mu_{HZ})^2 }   \cr
  && +\frac{ y_1(y_1-1) (\mu_Z -y_1)- y_2(1+y_1) (y_1+\mu_Z)+2 \mu_Z (1-4 \mu_H+\mu_Z)}{(y_1+\mu_{HZ})^2} \cr
  && +\frac{(y_1-1)^2 (\mu_Z-y_1)^2-\mu_Z^2+\mu_Z (1-4 \mu_H) (\mu_Z-4 \mu_H)}{4 (y_1+\mu_{HZ})^2 \mu_Z}  \cr
   && +\frac{ y_2(y_2-1) (\mu_Z-y_2)-y_1 (1+y_2) (y_2+\mu_Z)+2 \mu_Z (1-4 \mu_H+\mu_Z)}{(y_2+\mu_{HZ})^2} \cr
   && +\frac{(y_2-1)^2 (\mu_Z-y_2)^2-\mu_Z^2+ \mu_Z(1-4 \mu_H) (\mu_Z-4 \mu_H)}{4\mu_Z(y_2+\mu_{HZ})^2 }  \cr
   && +\frac{ y_1(y_1-1) (\mu_Z-y_1)-y_2(1+y_1) (y_1+\mu_Z)+2 \mu_Z (1-4 \mu_H+\mu_Z)}{(y_1+\mu_{HZ}) (y_2+\mu_{HZ})} \cr
   && +\frac{y_2(y_2-1) (\mu_Z-y_2)-y_1 (1+y_2) (y_2+\mu_Z)+2 \mu_Z (1-4 \mu_H+\mu_Z)}{(y_1+\mu_{HZ}) (y_2+\mu_{HZ})} \cr
   && +\frac{1}{2\mu_Z(y_1+\mu_{HZ}) (y_2+\mu_{HZ}) }\bigg\{\mu_Z^2+4 \mu_H \mu_Z (1+4 \mu_H+\mu_Z) \cr
   && +y_1 y_2 \left[1+y_1 y_2+\mu_Z^2+4 \mu_H (1+\mu_Z)\right] \cr
   && +(1+y_3 + 2 \mu_Z) [\mu_Z (y_3-8 \mu_H+\mu_Z)-y_1 y_2 (1+\mu_Z)]\bigg\} \ , \cr
{\cal Z}_3 &=& \frac{(y_1+y_2)^2}{4 \mu_Z}+2 \  , \cr
{\cal Z}_{12} &=& \frac{1}{(y_3-\mu_{HZ})(y_1+\mu_{HZ})}
\Bigg\{\mu_Z \left[(y_1+y_2)^2+8 \mu_Z\right] \cr
   && +\frac{1}{2} [ y_1(y_1-1) (\mu_Z-y_1)-y_2 (1+y_1) (y_1+\mu_Z)+2 \mu_Z (1-4 \mu_H+\mu_Z)]\Bigg\} \cr
&& + \frac{1}{(y_2+\mu_{HZ})(y_3-\mu_{HZ})}
\Bigg\{\mu_Z \left[(y_1+y_2)^2+8 \mu_Z\right] \cr
   && +\frac{1}{2} [ y_2(y_2-1) (\mu_Z-y_2)-y_1 (1+y_2) (y_2+\mu_Z)+2 \mu_Z (1-4 \mu_H+\mu_Z)]\Bigg\} \ , \cr
{\cal Z}_{13} &=& \frac{\left[(y_1+y_2)^2+8 \mu_Z\right]}{2(y_3-\mu_{HZ})} \ , \cr
{\cal Z}_{23} &=& \frac{(y_1+y_2)^2+8 \mu_Z}{y_1+\mu_{HZ}}
+ \frac{(y_1+y_2)^2+8 \mu_Z}{y_2+\mu_{HZ}} \cr
   && +\frac{ y_1 (y_1-1)(\mu_Z-y_1)- y_2(1+y_1) (y_1+\mu_Z)+2 \mu_Z (1-4 \mu_H+\mu_Z)}{2 (y_1+\mu_{HZ}) \mu_Z} \cr
   && +\frac{ y_2(y_2-1) (\mu_Z-y_2)-y_1 (1+y_2) (y_2+\mu_Z)+2 \mu_Z (1-4 \mu_H+\mu_Z)}{2 (y_2+\mu_{HZ}) \mu_Z} \ ,
\end{eqnarray}
where
$x_3 = 2 - x_1 -x_2$ is the scaled energy of the $Z$ boson,
$y_1 = 1 - x_1$, $y_2 = 1-x_2$,
$\mu_H = m^2_{S_1} /s$ is the square of the reduced mass of $S_1$, and
$\mu_{HZ} = \mu_H - \mu_Z$.

Then, we integrate the differential cross section to obtain the total production cross section
for the double Higgs-strahlung process of $S_1$,
$\sigma(ZS_1S_1)$ as a function of $m_{S_1}$.
The values of the relevant parameters are tha same as in Fig. 1, $10^5$ points in the parameter space.
The result is shown in Fig. 6, where we take $\sqrt{s} = 500$ GeV, the proposed
c.m. energy of the first stage of the ILC (ILC-500).

For the sake of comparison, we also plot the corresponding SM cross section in Fig. 5.
One may notice that most of the points are distributed close to the solid curve.
This implies that $S_1$ behaves very much alike the SM Higgs boson
with respect to the double Higgs-strahlung process.
Also, the absolute minimum of the production cross section of the lightest scalar Higgs boson
via the double Higgs-strahlung process is about 0.05 fb at the ILC-500.
In other words, the SUSY $E_6$ model predicts that $\sigma(ZS_1S_1)$ is larger than 0.05 fb,
whatever the parameter values.
Therefore, we expect that the ILC-500 would produce at least 5 $S_1$ events for the SUSY $E_6$ model
via the double Higgs-strahlung process,
if the ILC-500 has the integrated luminosity of 500 fb$^{-1}$ and the efficiency of 20 \%.

%*****************************************************************
\section{Conclusions}
%*****************************************************************

We have studied a supersymmetric $E_6$ model, which has two Higgs doublets and two Higgs singlets.
In particular, we have studied the production of the lightest scalar Higgs boson in $e^+e^-$ collisions,
via Higgs-strahlung process and double Higgs-strahlung process, for a reasonably established parameter space,
at the one-loop level by considering the contributions from top and scalar top quark loops.

At the one-loop level, the mass of the lightest scalar Higgs boson of the SUSY $E_6$ model
is estimated to be between 112 and 142 GeV.
Thus, its mass is comparable to the SM Higgs boson.
Not only its mass but also its production cross sections in $e^+e^-$ collisions
via Higgs-strahlung process and double Higgs-strahlung process are found
to be quite similar to those of the SM Higgs boson.
For the most part, this similarity may be attributed to the experimental constraints
on the large masses of the extra neutral
gauge bosons and the tiny mixings between them and the $Z$ boson,
the electroweak precision measurement at the LEP2 experiment
and the direct search in $p{\bar p}$ collisions at the Tevatron.
Since the extra neutral gauge bosons are heavier than 800 GeV and their mixing angles with the $Z$ boson are less than
$3 \times 10^{-3}$, they are practically decoupled from $Z$ boson, and hence from $S_1$ in the
Higgs-strahlung process and double Higgs-strahlung process.
In other words, we could assume without difficulty that only the $Z$ boson is involved in these processes.
We also find that the same experimental constraints set the lower bound on $x_1$ and $x_2$,
the vacuum expectation values of
two Higgs singlets, as about 1265 GeV.

Our study on the $G_{ZZS_1}$ coupling coefficient, the production cross section $\sigma (e^+e^- \rightarrow Z \rightarrow ZZS_1)$ via
Higgs-strahlung process, the $G_{S_1 S_1 S_1}$  cubic coupling coefficient, $G_{ZZ S_1 S_1}$ quartic coupling coefficient,
and the production cross section $\sigma(e^+e^- \rightarrow Z \rightarrow Z S_1 S_1)$ via double Higgs-strahlung process
show clearly the similarity of $S_1$ of our model with the SM Higgs boson.
Thus, the contributions of heavier scalar Higgs bosons are nearly negligible in these processes.
Consequently, the experimental constraints on the extra neutral gauge bosons make the Higgs sector of the supersymmetric $E_6$ model with
two Higgs doublets and two Higgs singlets very similar to the SM Higgs sector, as far as $S_1$ is concerned.

The absolute lower bound on the cross section for $S_1$ production in $e^+e^-$ collisions via Higgs-strahlung process at the ILC-500
is about 19 fb, and
the absolute lower bound on the cross section for $S_1 S_1$ production in $e^+e^-$ collisions via double Higgs-strahlung process
at the ILC-500 is about 0.05 fb.
Assuming  the integrated luminosity of 500 fb$^{-1}$ and the efficiency of 20 \% for the ILC-500,
at least 5 events of $S_1$ of our model might be explored at the ILC via the double Higgs-strahlung process.

%*****************************************************************
\section{Acknowledgments}
%*****************************************************************

S. W. Ham thanks P. Ko for the hospitality at KIAS where a part of this work has been performed.
He thanks Kihyeon Cho at High Energy Physics Team, KISTI for the collaboration.
This research was supported by Basic Science Research Program
through the National Research Foundation of Korea (NRF) funded
by the Ministry of Education, Science and Technology
(2009-0086961, 2009-0070667).

%******************************************************************

\vfil\eject
%%%%%%%%%%%%%%%%%%%%%%%%%%%%%%%%%%%%%%%%%%%%%%%%%%%%%%%%%%%%%%%%%%%%%%%%

%Figure
\setcounter{figure}{0}
\def\figurename{}{}%
% (FIG 1)
\renewcommand\thefigure{FIG. 1}
\begin{figure}[t]
\begin{center}
\includegraphics[scale=0.6]{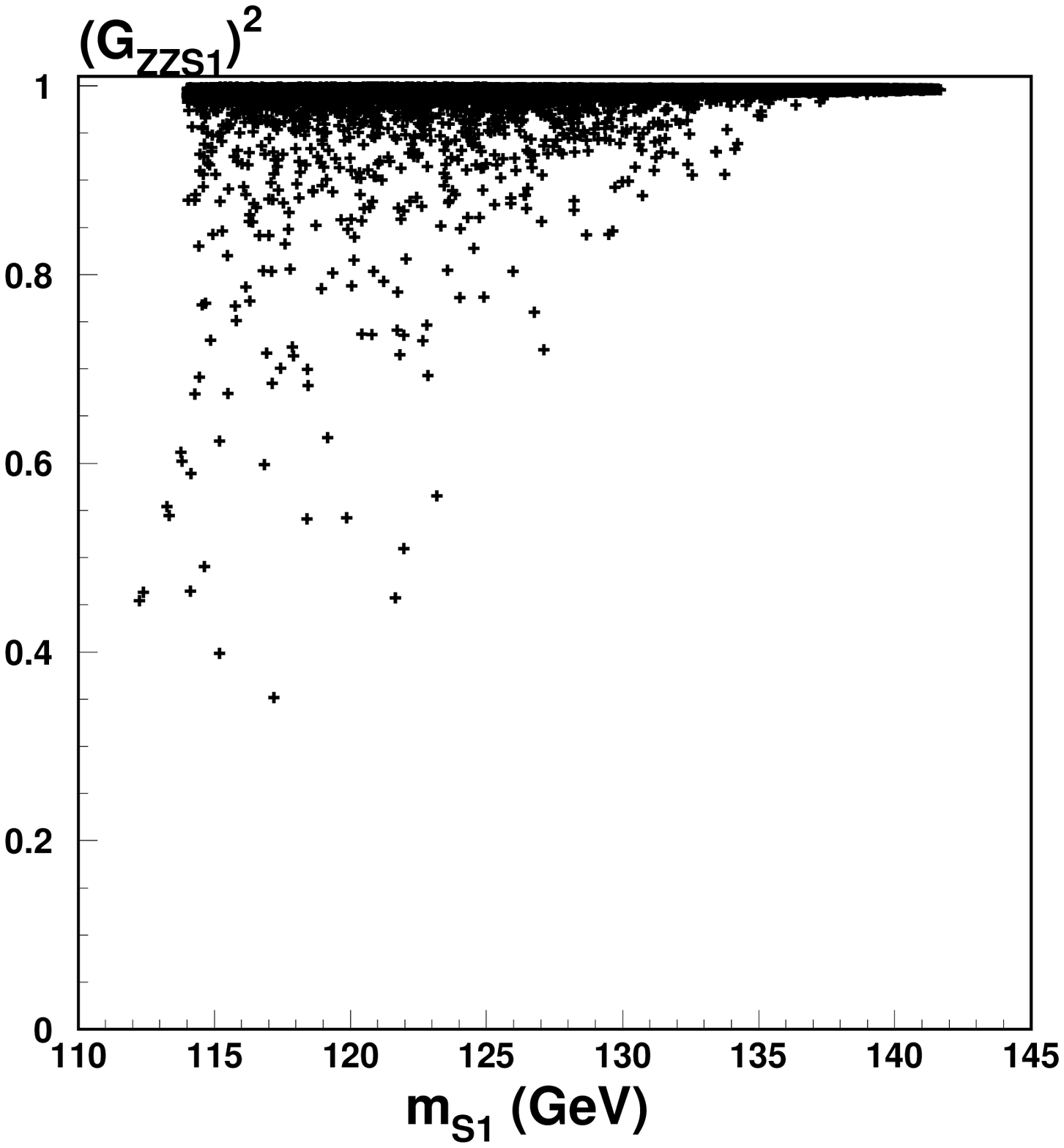}
\caption[plot]{The squared coupling coefficient $G_{ZZS_1}^2$ against $m_{S_1}$,
normalized by the SM coefficient, for $10^5$ sets of parameter values
that vary within the parameter space established as $1 < \tan \beta \leq 30$,
$0 < \lambda \leq 0.83$, $ 10 < A_{\lambda} < 400$,
$0  < \theta < \pi/2$, $100 \leq x_1, \ x_2 \leq 1500$ GeV,
$100 \leq m_Q, \ m_T, \ A_t \leq 1000$ GeV.
The squared SM coefficient, normalized by itself, is a solid line at 1.0.
Notice that most of the points are distributed close to the solid line,
and that $m_{S_1}$ is between 112 and 142 GeV.}
\end{center}
\end{figure}

%Figure
\setcounter{figure}{0}
\def\figurename{}{}%
% (FIG 2)
\renewcommand\thefigure{FIG. 2}
\begin{figure}[t]
\begin{center}
\includegraphics[scale=0.6]{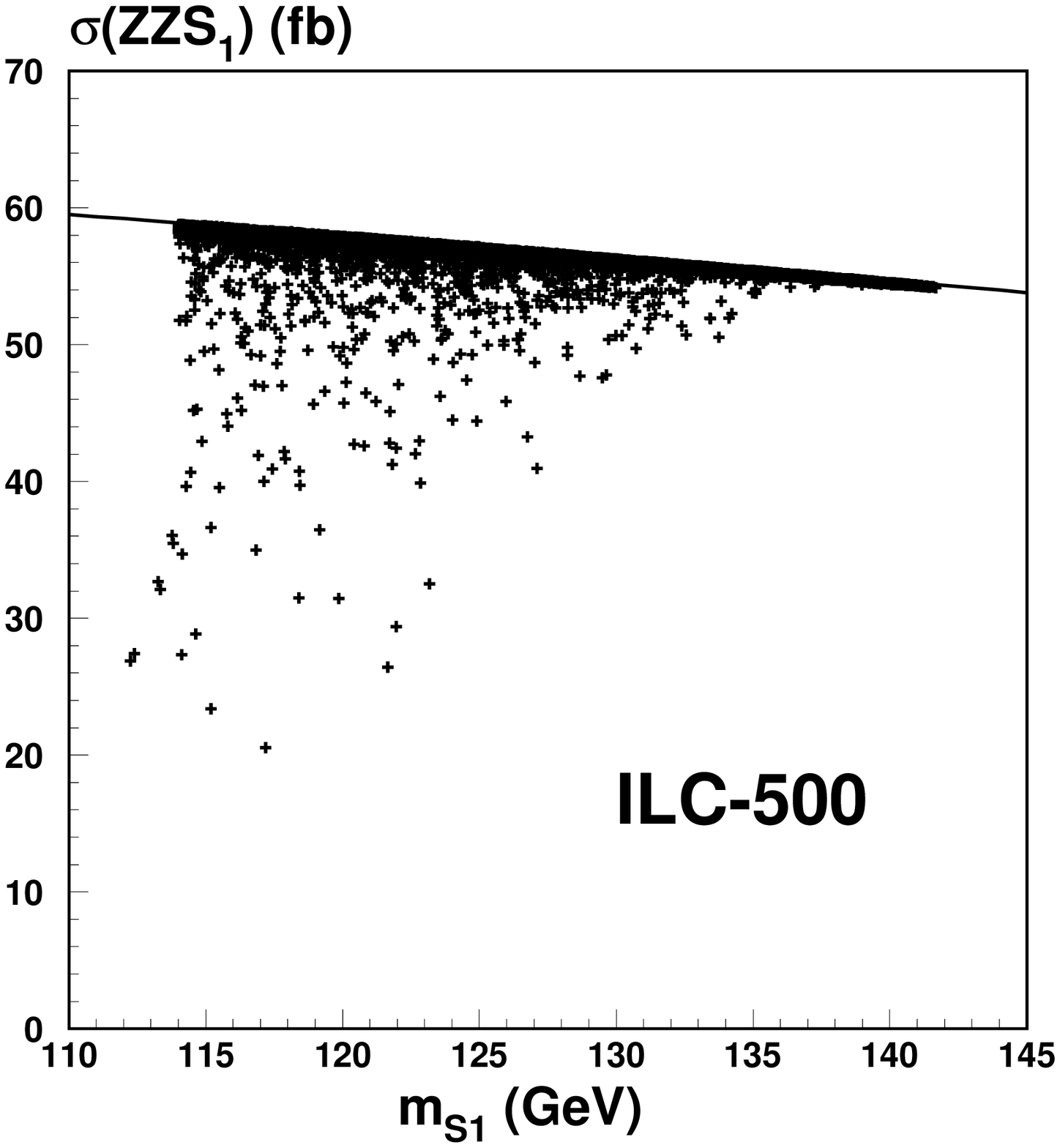}
\caption[plot]{The production cross section $\sigma(ZZS_1)$ against $m_{S_1}$, via Higgs-strahlung process in
$e^+ e^-$ collisions with $\sqrt{s} = 500$ GeV, for the same $10^5$ sets of parameter values as in Fig. 1.
The solid curve is the corresponding SM cross section as a function of the SM Higgs boson mass.
}
\end{center}
\end{figure}

%%%%%%
% (FIG 3)
\renewcommand\thefigure{FIG. 3}
\begin{figure}[t]
\begin{center}
\includegraphics[scale=0.8]{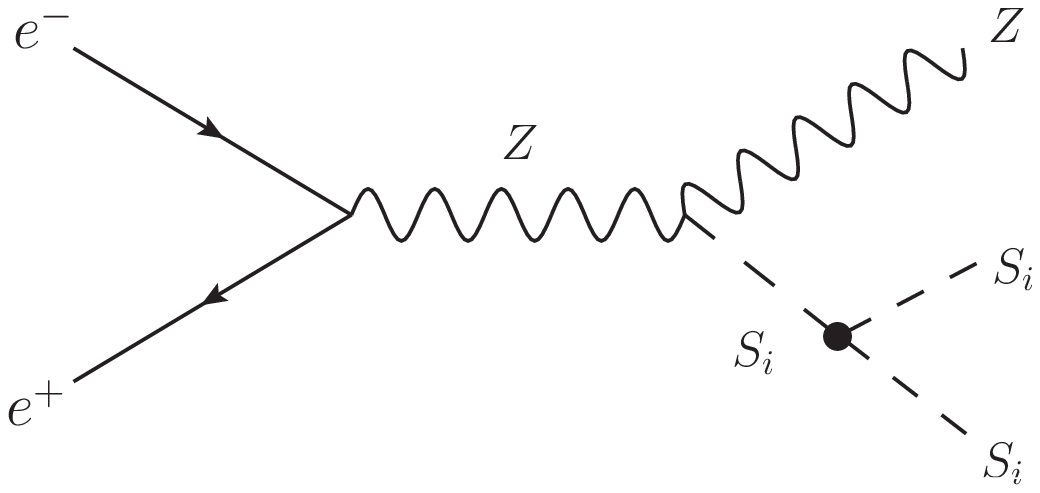}
\includegraphics[scale=0.8]{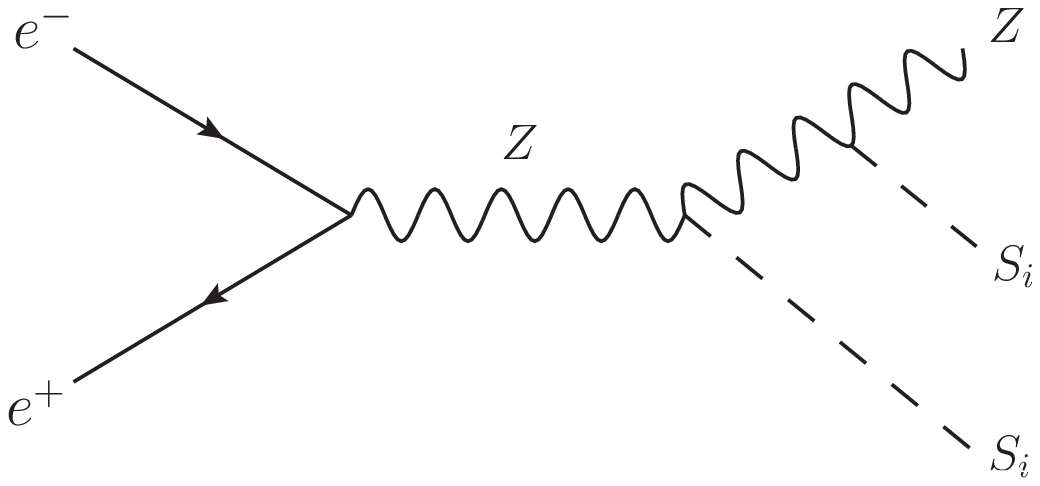}
\includegraphics[scale=0.8]{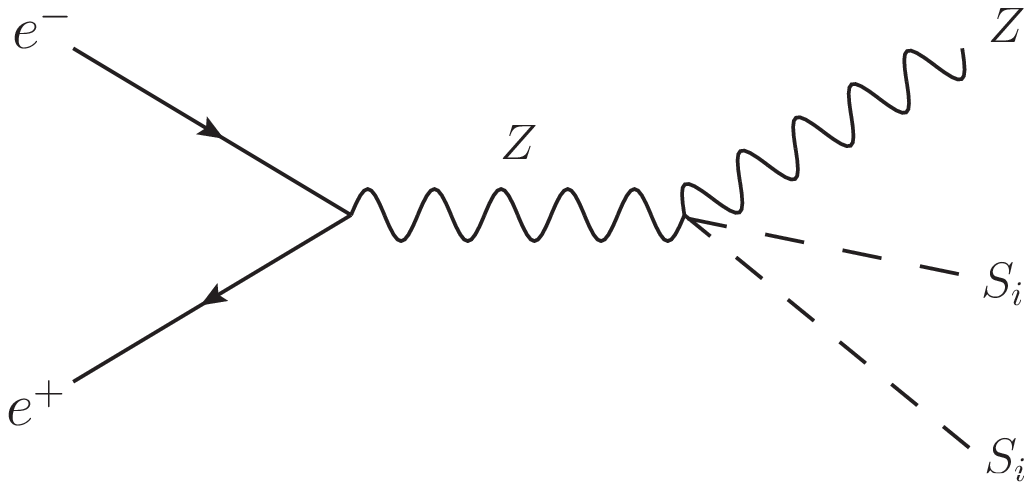}
\caption[plot]{Feynman diagrams for double Higgs-strahlung process, $e^+e^- \rightarrow Z \rightarrow ZS_1S_1$.
Notice that the Higgs cubic coupling and the Higgs-Higgs-gauge-gauge quartic coupling are present.}
\end{center}
\end{figure}
%%%%%%

%Figure
\setcounter{figure}{0}
\def\figurename{}{}%
% (FIG 4)
\renewcommand\thefigure{FIG. 4}
\begin{figure}[t]
\begin{center}
\includegraphics[scale=0.6]{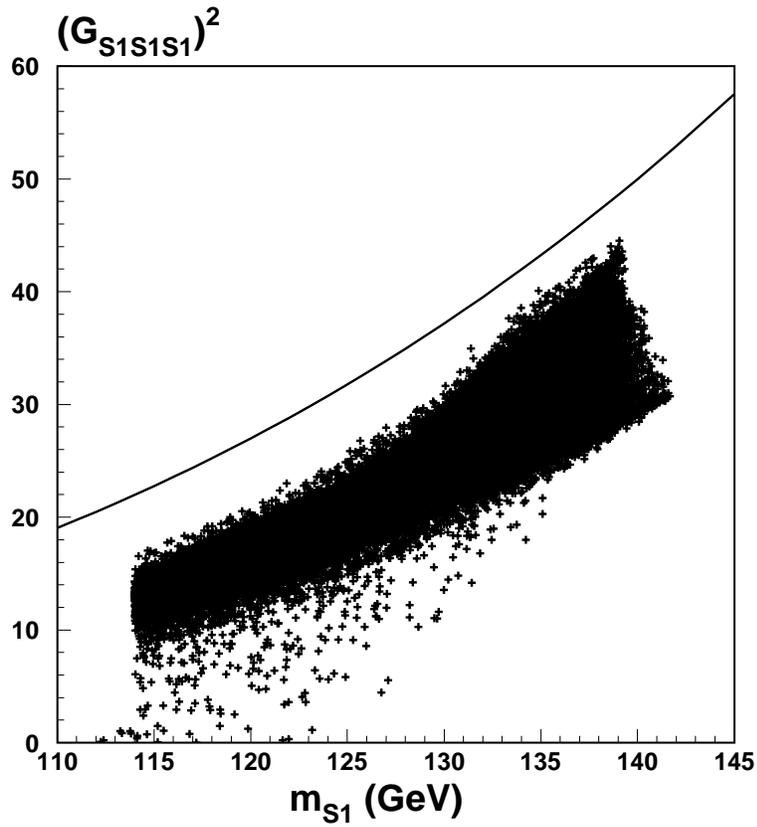}
\caption[plot]{The squared coupling coefficient $G_{S_1 S_1 S_1}^2$ against $m_{S_1}$,
normalized by $2 m_Z^2/v $, for same $10^5$ sets of parameter values as in Fig. 1.
The solid curve is the squared SM coefficient, normalized by $2 m_Z^2/v $, as a function of the SM Higgs boson mass.
}
\end{center}
\end{figure}

% (FIG 5)
\renewcommand\thefigure{FIG. 5}
\begin{figure}[t]
\begin{center}
\includegraphics[scale=0.6]{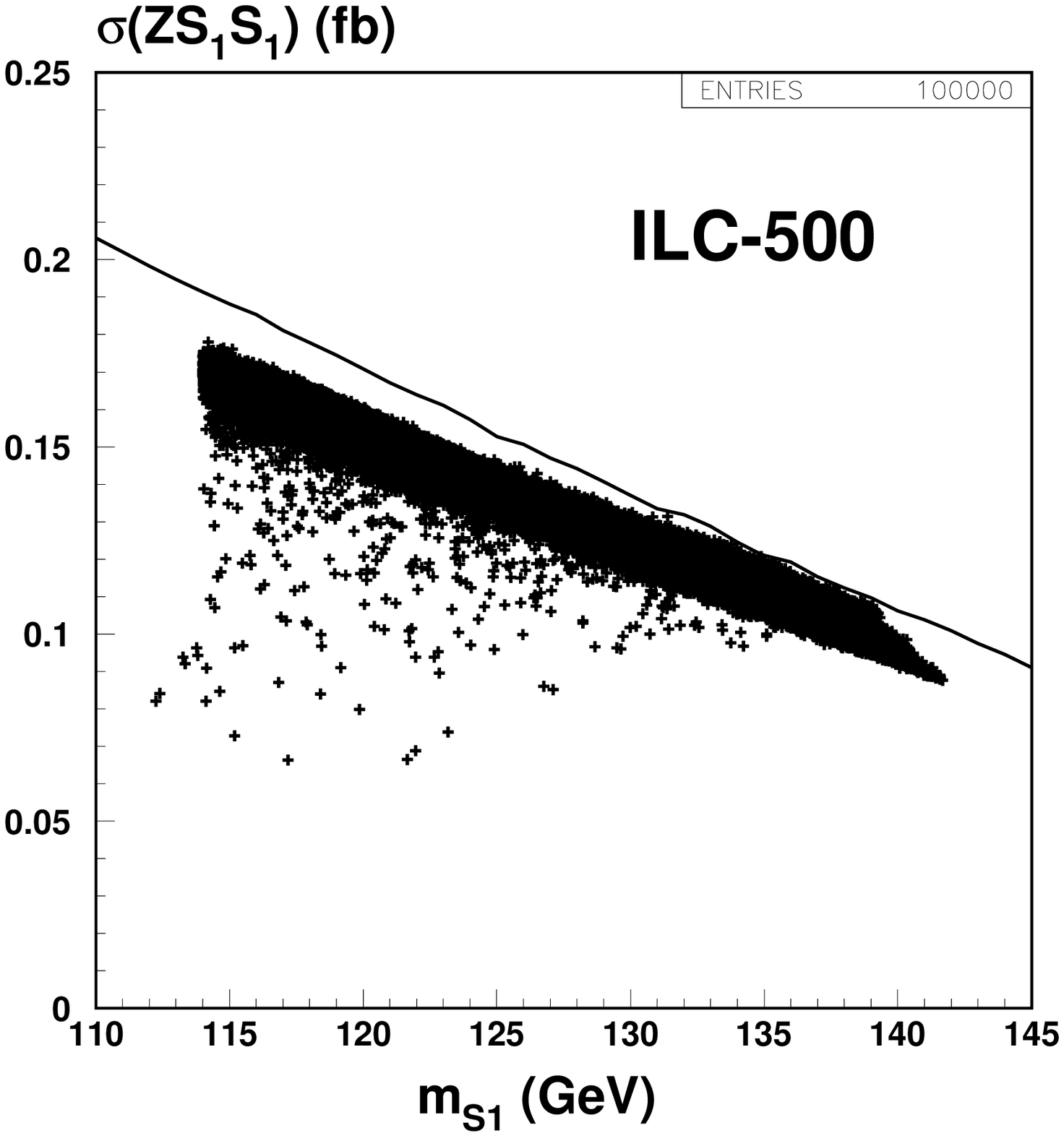}
\caption[plot]{The production cross section $\sigma(ZS_1S_1)$ against $m_{S_1}$, via double Higgs-strahlung process in
$e^+ e^-$ collisions with $\sqrt{s} = 500$ GeV, for the same $10^5$ sets of parameter values as in Fig. 1.
The solid curve is the corresponding SM cross section as a function of the SM Higgs boson mass.
}
\end{center}
\end{figure}

%***********************************************************************

\begin{thebibliography}{99}
%******************************************************************
\bibitem{1} P. Fayet and S. Ferrara, Phys. Rep. {\bf 32}, 249 (1977); P. Fayet, Phys. Rep. {\bf 105}, 21 (1984).
\bibitem{2} L. Girardello and M. T. Grisaru, Nucl. Phys. B {\bf 194}, 65 (1982).
\bibitem{3} H. P. Nilles, Phys. Rep. {\bf 110}, 1 (1984).
\bibitem{4} J. F. Gunion, H. E. Haber, G. L. Kane, and S. Dawson, {\it The Higgs Hunters' Guide}
(Addison-Wesley, CA, 1990).
\bibitem{5} P. Fayet, Nucl. Phys. {\bf 90}, 104 (1975);
P. Fayet, Phys. Lett. B {\bf 69}, 489 (1977);
P. Fayet, Phys. Lett. B {\bf 125}, 178 (1983);
E. Cremmer, P. Fayet, and L. Girardello, Phys. Lett. B {\bf 122}, 41 (1983).
\bibitem{6} J. Ellis, J. F. Gunion, H. E. Haber, L. Roszkowski, F. Zwirner, Phys. Rev. D {\bf 39}, 844 (1989).
\bibitem{7} J. E. Kim and H. P. Nilles, Phys. Lett. B {\bf 138}, 150 (1984).
\bibitem{8} H. E. Haber and M. Sher, Phys. Rev. D {\bf 35}, 2206 (1987).
\bibitem{9} J. L. Hewett and T. G. Rizzo, Phys. Rep. {\bf 183}, 193 (1989).
\bibitem{10} J. F. Gunion, L. Roszkowski, and H. E. Haber, Phys. Lett. B {\bf 189}, 409 (1987);
J. F. Gunion, L. Roszkowski, and H. E. Haber, Phys. Rev. D {\bf 38}, 105 (1988).
\bibitem{11} S. W. Ham, J. O. Im, E. J. Yoo, and S. K. Oh, JHEP 12, 017 (2008).
\bibitem{12} A. Djouadi, H. E. Haber, and P. M. Zerwas, Phys. Lett. B {\bf 375}, 203 (1996).
\bibitem{13} A. Djouadi, W. Kilian, M. M\"{u}hlleitner, and P. M. Zerwas, Eur. Phys. J. C {\bf 10}, 27 (1999).
\bibitem{14} The LEP Collaborations ALEPH, DELPHI, L3 and OPAL,
{\it Search for neutral MSSM Higgs bosons at LEP}, Eur. Phys. J. C {\bf 47}, 547 (2006).
\bibitem{15} C. Amsler {\it et al}. (The Particle Data Group), Phys. Lett. B 667, 1 (2008).
\bibitem{16} S. Coleman and E. Weinberg, Phys. Rev. D {\bf 7}, 1888 (1973).
\bibitem{17} S. W. Ham, S. K. Oh, D. Son, Phys. Rev. D {\bf 65}, 075004 (2002).
\bibitem{18} P. Osland and P. N. Pandita, Phys. Rev. D {\bf 59}, 055013 (1999).
\bibitem{19} A. Arhrib, R. Benbrik, and C. W. Chiang, Phys. Rev. D {\bf 77}, 115013 (2008).
\end{thebibliography}
\end{document}